\documentstyle[epsfig,psfig]{aipproc2}
\newcommand{\epem}{\mbox{$e^+e^-$}}
\newcommand{\Emiss}{\mbox{$\protect \raisebox{.3ex}{$\not$}{\rm E}$}}
\begin{document}
\title{Physics at \epem~Linear Colliders}

\author{David W. Gerdes}
\address{University of Michigan, Department of Physics \\ 
         2477 Randall Laboratory, 500 E. University Ave. \\
         Ann Arbor, MI 48109-1120 USA \\
         {\em E-mail: gerdes@umich.edu} \\ \vspace*{0.2in}
{\em To appear in the proceedings of the 5th International Conference on
   Physics Potential and Development of $\mu^+\mu^-$ Colliders (MUMU99), 
    December 15-17, 1999, 
    San Francisco, CA}
        }%
\maketitle
\begin{abstract}
I discuss the motivation and physics potential of an electron-positron linear 
collider with a center-of-mass energy at the 1 TeV scale, in light of what
we may expect to learn with the LHC. The comparison is illustrated
with examples drawn from Higgs physics, top quark physics, and the search for 
large extra spacetime dimensions. 

\end{abstract}

\section*{Introduction}
The past decade of precision electroweak experiments has seen outstanding
confirmation of the Standard Model at the per mille level. But the 
successes of the Standard Model have drawn increased attention to its
deficiencies, notably its unsatisfactory treatment of the mechanism behind
electroweak symmetry breaking. This phenomenon occurs roughly at the 
1~TeV scale, which the LHC will access directly. The merits of any additional 
machine must therefore be evaluated within the context of the LHC program. 
A TeV-scale \epem~linear collider (LC) is a leading candidate for such a
facility. Such a machine offers control over the
beam energy and polarization, and a clean environment that enables precision 
event reconstruction.  In this paper I illustrate aspects of the 
LC physics program with 
examples drawn from Higgs physics, top quark physics, and the study
of large extra spacetime dimensions. More comprehensive reviews
of physics at $e^+e^-$ linear colliders are given in Refs.~\cite{Snowmass96},
 \cite{Peskin-Murayama}, and \cite{TESLA-review}.

\section*{Machine and Detector Overview}
Well-developed LC designs have been put forward by the SLAC-KEK joint effort 
(the NLC/JLC designs)\cite{NLC-ZDR} and by DESY (the TESLA 
design)\cite{TESLA-design}. The two machines differ 
technologically but achieve similar ends. The NLC 
uses warm rf cavities operating in the X-band (11.4 GHz). 
The baseline design assumes initial operation at 
$\sqrt{s}=500$~GeV at a luminosity of $5\times 10^{33}$~cm$^{-2}$s$^{-1}$, with
an $e^-$ beam polarization of 80-90\%. The 
linac is designed to allow adiabatic energy upgrades to $\sqrt{s}=1$~TeV
through the addition of klystrons. The size of the beam-delivery and final 
focus systems would allow eventual operation at 1.5 TeV. The TESLA design
uses superconducting rf operating at $1.3$~GHz, and reaches a center of 
mass energy of 800-1000~GeV. Initial operation would be in the 
200-500 GeV range,
with an $e^-$ polarization of 80\% as well as a positron beam polarization of
60\%, which would introduce new measureables into physics processes.
This design permits operation at very high luminosity,
up to $5\times 10^{34}$~cm$^{-2}$s$^{-1}$. The two designs 
have rather
different beam characteristics and time structures. A brief comparison of the 
TESLA and NLC parameters is shown in Table~\ref{tab:machines}; a more complete
list can be found in Ref.~\cite{machine-params}.

Designing a detector for a linear collider is widely regarded, 
at least by those who work at hadron colliders, as an easy problem. 
Certainly the LC does not share the LHC's formidable challenges of high 
event rates and radiation exposures. The challenges for a linear collider 
detector stem from the desire to fully exploit the clean machine 
environment
by building a detector of the highest possible precision. Current designs
are evolutionary extensions of the LEP and SLD detectors and feature
CCD pixel vertex detectors, silicon or TPC outer trackers, and a fine-grained
EM calorimeter located inside the magnet coil. Several designs incorporate
the hadronic calorimeter inside the coil as well. In contrast to the LHC where
triggering is a major challenge, at a linear collider the full detector can 
be read out between bunch trains and triggers formed in software.

\begin{table}[ht]
    \caption{Summary of JLC/NLC and TESLA accelerator parameters. Luminosities
 include the pinch enhancement.}
    \label{tab:machines}
    \leavevmode
    \begin{tabular}{lllllll}
     Design &  $\sqrt{s}$~(GeV) & Lum.  & bunches/ & bunch & $\delta E/E$&  $\sigma_x/\sigma_y$\\ 
    & &($\times 10^{33}$) & pulse &sep. (ns) & (\%) & at IP (nm) \\ \hline
     JLC/NLC 500 & 500 & 6.5 & 95 & 2.8 & 3.7 & 330/4.9 \\ 
     JLC/NLC 1000 & 1000 & 12.9 & 95 & 2.8 & 10.3 & 234/3.9\\
     JLC/NLC 1500(B) & 1478 & 12.4 & 95 & 2.8 & 14.1 & 200/3.7 \\
     TESLA & 500 & 30 & 2820 & 337 & 2.8 & 553/5 \\
     TESLA & 800 & 50 & 4500 & 189 & 4.7 & 391/2\\ \hline\hline
    \end{tabular}
\end{table}

\section*{Light Higgs Physics}

Current electroweak data point to the 
existence of a light Higgs between roughly 100 and 200 GeV, with
the lower end of this range being favored by the fits\cite{LEP-EWWG}.
If so, the Higgs may be discovered in the near future at LEP\cite{Wu} or
the Tevatron\cite{Hobbs}, but if not there then certainly at the LHC. A Higgs
in this mass range can be convincingly observed at the LHC through such channels
as $H\rightarrow \gamma\gamma$, $H\rightarrow ZZ^{(*)}$ or $WW^{(*)}$, and
production of $t\bar{t}H$ followed by $H\rightarrow \gamma\gamma$ or 
$b\bar{b}$\cite{ATLAS-phyTDR}. Yet the LHC cannot see, or can see only with
great difficulty, many important Higgs decays, such as $H\rightarrow c\bar{c}$
and $H\rightarrow\tau^+\tau^-$, that are critical to determining if this object
is indeed \textit{the} Higgs, the relic of electroweak symmetry breaking that 
couples to fermions in proportion to their masses. 

\begin{figure}[t]
\centerline{
\vspace*{-0.2cm}
\psfig{figure=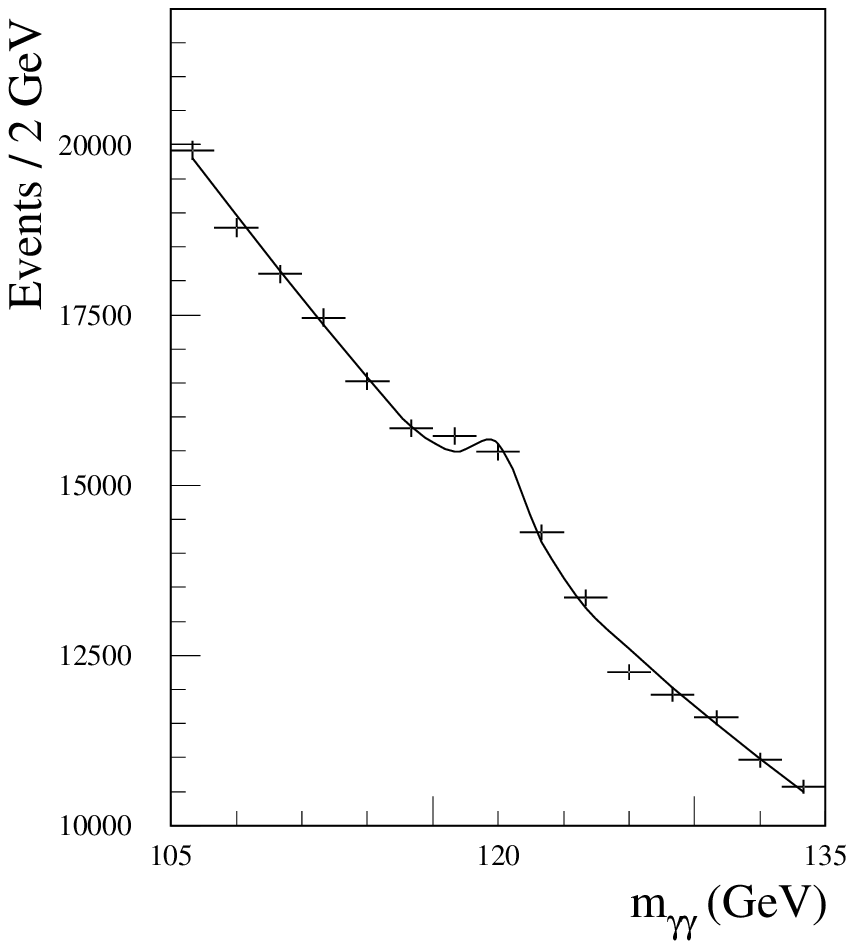,height=8.cm,width=8cm,angle=0}
\hspace*{0mm}\vspace*{0.2cm}
\psfig{figure=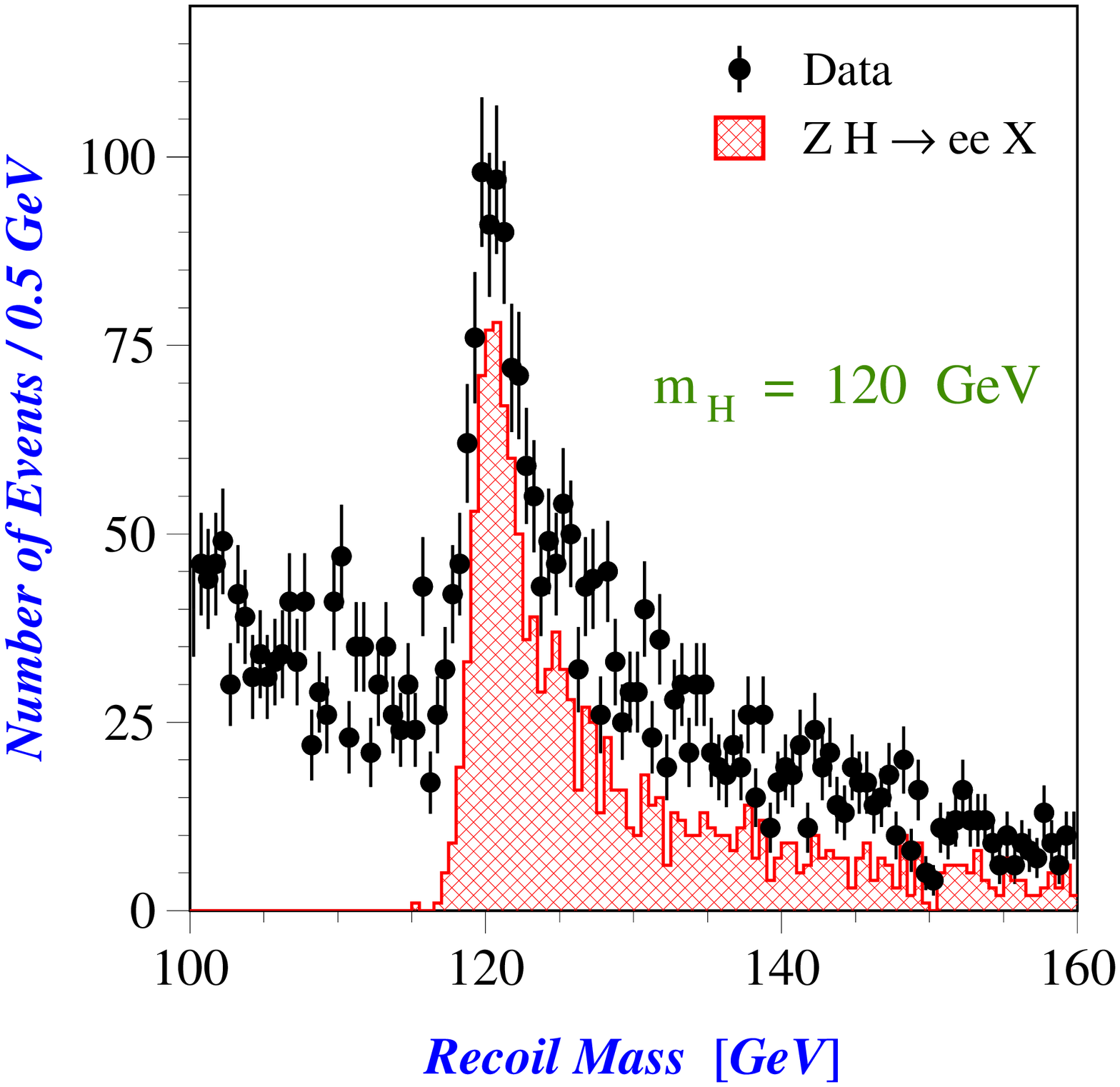,height=8.5cm,width=8.5cm,angle=0}}
\vspace*{0.2cm}
\caption{120 GeV Higgs comparison. 
Left: $H\rightarrow\gamma\gamma$ signal in 100 fb$^{-1}$ at the 
LHC\protect\cite{ATLAS-phyTDR}. Right: $ZH\rightarrow\ell^+\ell^-X$ 
signal at the LC for 500 fb$^{-1}$
at $\sqrt{s}=350$ GeV\protect\cite{Higgs-recoil}.}
\label{fig:Higgs-120}
\end{figure}

At the LC, a light Higgs can be cleanly observed in recoil off the 
$Z$\cite{Higgs-recoil,juste}.
The cross section for this process peaks in the 250-400 GeV range, making light
Higgs physics an attractive target for the initial phase of the LC physics
program. Since the signature of this process is a monoenergetic $Z$ boson, 
these events can be reconstructed with high efficiency independent of the 
Higgs decay mode. This gives a clean inclusive sample in which to study Higgs 
decays, measure $m_H$ to 100-200 MeV (similar to the LHC), and obtain an 
extremely precise
measurement of the $H$-$Z$ Yukawa coupling\cite{juste}.
A sample recoil mass plot is shown in Figure~\ref{fig:Higgs-120}, in
comparison to the dominant $H\rightarrow\gamma\gamma$ discovery signal for the
same-mass Higgs at the LHC.

\begin{figure}[htb]
  \begin{center}
    \leavevmode
    \epsfysize=3.0in
    \epsffile{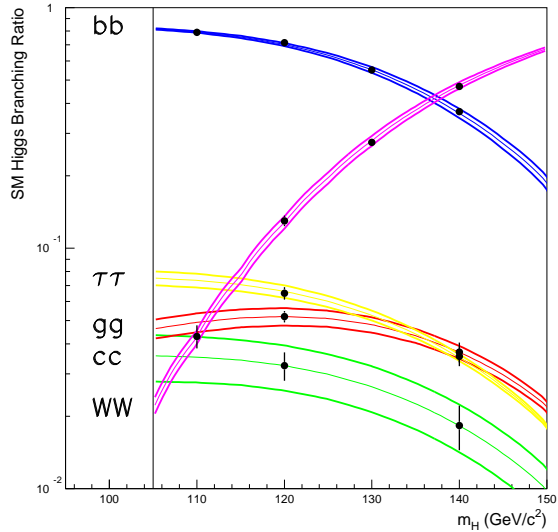}
    \caption{Predicted SM Higgs branching ratios, together with the measured
   values obtainable with 500~fb$^{-1}$ at $\sqrt{s}=350$ GeV, from the study
   in Ref.~\protect\cite{Battaglia-sitges}.}
    \label{fig:Higgs-BR}
  \end{center}
\end{figure}

To exploit this inclusive sample fully, however, it is necessary to have a
vertex detector capable of cleanly separating bottom, charm, and light quark
jets. This ability is provided at the LC by a CCD pixel vertex detector, which
can be located as close as 1~cm from the beam. Such a device is far too slow
and rad-soft to be practical at a hadron collider, but the more forgiving
environment of the LC allows one to exploit its superior spatial resolution
for excellent flavor separation. The payoff of this capability is demonstrated
in a recent study by Battaglia\cite{Battaglia-sitges}, summarized in
Figure~\ref{fig:Higgs-BR}. With 500~fb$^{-1}$ at $\sqrt{s}=350$~GeV, the
branching ratios $H\rightarrow b\bar{b}, c\bar{c}, \tau^+\tau^-, 
gg,~\mathrm{and}~WW^*$ can be measured with an accuracy of 2-5\%. Interestingly,
the $H\rightarrow\gamma\gamma$ mode, which is the prime discovery channel for
a $\sim 120$~GeV Higgs at the LHC, is undetectable in this production 
mode at the LC due to its very small ($\sim 10^{-3}$) branching fraction.
(The inverse process $\gamma\gamma\rightarrow H$ can be observed if the
LC is operated as a $\gamma\gamma$ collider, using backscattered Compton
photons from the primary $e^+e^-$ beams.)

This ensemble of branching ratio measurements can be used to 
distinguish a SM Higgs from the lightest Higgs ($h^0$) of 
the MSSM. Typically the branching ratios of the $h^0$ are equal to those
of the SM Higgs, times a function of $\tan\beta$ and $M_{A^0}$, where the 
$A^0$ is the heavy, CP-odd Higgs of the MSSM. A likelihood fit can therefore 
be used to determine whether the collection of observed BR's is more 
consistent with the SM or with the MSSM. Separation of the SM from the MSSM
Higgs can be determined at the 90\% confidence level with the above 
measurements for $M_{A^0}$ up to 730 GeV, with the dominant uncertainty 
coming from knowledge of the $b$ and $c$ quark masses\cite{Battaglia-sitges}.
If SUSY exists at this scale, it will most likely have been discovered at
the LHC, but measurements such as this will constitute a vital precision test
that can only be performed at the LC, as a muon collider, too, has difficulties
with high-precision charm ID.

\section*{Top Quark Physics}

The top quark's privileged status as the most massive known matter particle, 
and the only fermion with a mass at the ``natural'' electroweak scale, make
it a prime target for all future colliders. The LC aims to carry out a complete
program of top quark physics, including measurements of top's mass, width,
form factors, and, perhaps most interestingly, its Yukawa coupling to the
Higgs. Furthermore, the process $\epem\rightarrow t\bar{t}\nu\bar{\nu}$, 
accessible at a 1.5~TeV LC, can be a sensitive probe of electroweak 
symmetry-breaking by new strong interactions\cite{ttbarnunu}.

The mass of the top quark, $m_t$, is a precision electroweak parameter that affects
relationships among other electroweak observables such as $M_W$, $M_Z$,
$\sin^2\theta_W$, and $m_H$. Future measurements at the Tevatron and the
LHC are likely to give a 2-3 GeV precision on $m_t$, dominated by systematics.
At the LC, the top quark's mass can be determined to about 100-200 MeV, and the 
width to about 7\%, in a relatively low-luminosity (10-50 fb$^{-1}$) threshold 
scan\cite{top-threshold}. But what would we gain from such a high precision 
measurement? Table~\ref{tab:mtop-mhiggs} shows the fractional precision on
$m_H$ that would follow from various uncertainties on $M_W$ and
$m_t$\cite{TESLA-review}. A 200~MeV
uncertainty on $m_t$, together with a 15~MeV uncertainty on $M_W$ (which may be
achievable from a high-luminosity return to the $W$ pair threshold with the 
LC) yields a 17\% uncertainty on $m_H$. For comparison, an uncertainty of
about 50\% is expected from measurements at LEP~II and the Tevatron. The Higgs
is likely to have been discovered by the time the LC makes this measurement,
in which case it will serve as a key consistency test---much like the comparison
of the directly measured $m_t$ to the value inferred from electroweak data does 
today.
\begin{table}[htbp]
  \begin{center}
    \caption{Fractional uncertainty on the Higgs mass for various 
     uncertainties on $M_W$ and $m_t$, from Ref.~\protect\cite{TESLA-review}.
     A 30~MeV uncertainty on $M_W$ would seem more appropriate for LEP~II+LHC
     and would yield a larger uncertainty on $m_H$.}
    \label{tab:mtop-mhiggs}
    \leavevmode
    \begin{tabular}{lll}
      Experiment & $\delta M_W, \delta m_t$ & $\delta m_H/m_H$ \\ \hline
      LEP~II + Tevatron & 30~MeV, 4 GeV & 57\% \\
      LEP~II + LHC & 15~MeV, 2~GeV & 26\% \\
      LC   & 15 MeV, 200 MeV & 17\% \\ \hline\hline
    \end{tabular}
  \end{center}
\end{table}

Of still greater interest, however, is a direct measurement of the top-Higgs
Yukawa coupling, $\lambda_{t\bar{t}H}$. 
Such a measurement, like those of the Higgs branching
ratios discussed above, is needed to establish the ``Higgsness'' of the Higgs,
and may also probe the special nature of the top quark. At the LHC, the
ratio $\lambda_{t\bar{t}H}/\lambda_{WH}$ can be measured to an accuracy of 25\%
for $80<m_H<120$~GeV\cite{ATLAS-phyTDR}.
For a light Higgs, $\lambda_{t\bar{t}H}$ can be measured at the LC in $t\bar{t}H$
production. For Higgs masses around 120~GeV, the cross section for this process
peaks at about 2.6~fb for $\sqrt{s}=700$-$800$~GeV and then falls off slowly.
This is some three orders of magnitude smaller than the dominant $t\bar{t}$,
$WW$, and $t\bar{t}Z$ backgrounds. But the spectacular nature of these
events ($qqqqbbbb$ if both tops decay hadronically, or $qqbbbb + \ell^{\pm}$ + 
\Emiss~if both tops decay semi-leptonically), and their many kinematic
constraints, provide enough handles that backgrounds can be acceptably 
reduced through direct mass reconstruction\cite{baer} or a
neural net\cite{juste2}. In the latter study, the authors assume 1000~fb$^{-1}$
(about 3 years of running at ${\cal L} = 10^{34}$) at $\sqrt{s}=800$~GeV, and
obtain a 5.5\% uncertainty on $\lambda_{t\bar{t}H}$. Outstanding flavor-ID is again
a prerequisite for this measurement. 
The possibility of measuring 
$\lambda_{t\bar{t}H}$ with such high precision is a strong argument in favor of 
the highest possible luminosities.

\section*{Large Extra Dimensions}

The recent proposal\cite{lowscale,ADD} to resolve
the hierarchy problem through a theory of low-scale quantum gravity with large 
extra spacetime dimensions has generated great interest because
of its testable consequences at colliders\cite{ADD-pheno,Leff,Hewett}. In
these models, Standard Model fields are confined to the 4-dimensional boundary
of a ``bulk'' with $n$ compact extra dimensions of characteristic size $R$. 
Gravitons propagate in the bulk, where they couple with a strength of order the 
electroweak strength (hence the elimination of the hierarchy). The apparent 
weakness of gravity in our 4-dimensional world arises from the geometrical
suppression of the gravitational flux lines by a factor proportional to the
volume of the compact extra dimensions:
$$
                 M_{\mathrm{Pl}}^2 = V_n M_s^{n+2},
$$
where $M_{\mathrm{Pl}}=10^{19}$~GeV is the Planck scale, $V_n\sim R^n$ is
the volume of the compact extra dimensions, and $M_s$ is the fundamental 
Planck scale in the bulk. If we require $M_s = {\cal{O}}(\mathrm{few~TeV})$
to eliminate the hierarchy, we can obtain the characteristic size $R$ of the 
extra dimensions for various values of $n$. Values of $R$ as large as a fraction 
of a millimeter are permitted by current limits from Cavendish-type 
experiments\cite{long}. 

For phenomenological purposes, we are most concerned with the effective Lagrangian
that describes the interactions between gravitons and SM fields in our 4-dimensional
world\cite{Leff}. Gravitons then appear as a Kaluza-Klein tower, or series, of 
closely-spaced massive spin-2 states that can be emitted or exchanged along with 
SM gauge bosons.
Each such state has a very weak coupling to matter, of order $1/M_{\mathrm{Pl}}$,
but because of the large number of these states their cumulative effect is comparable
to that of Standard Model processes at energies near $M_s$. 

\begin{figure}[t]
\centerline{
\psfig{figure=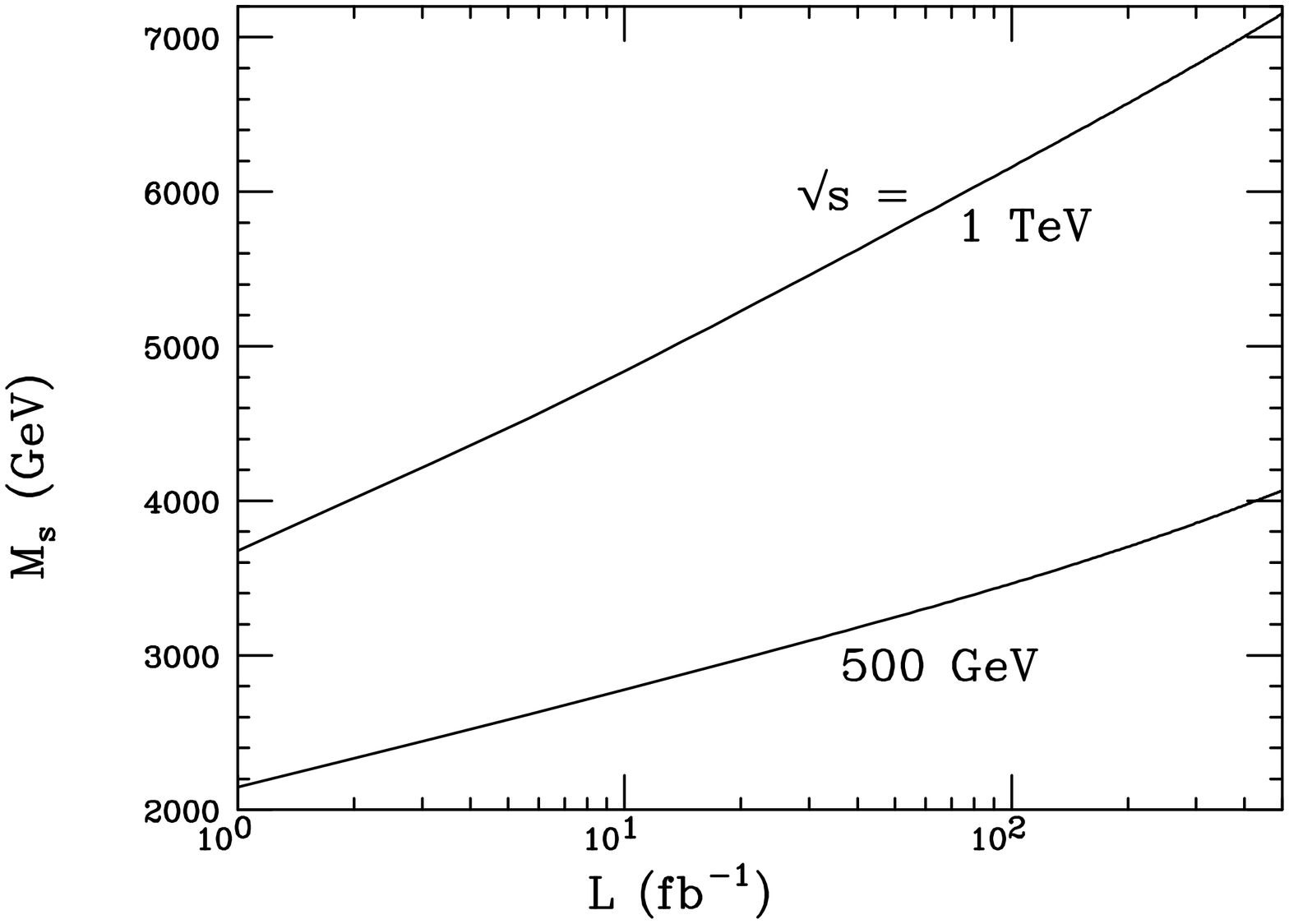,height=8.cm,width=8cm,angle=-90}
\hspace*{-5mm}
\psfig{figure=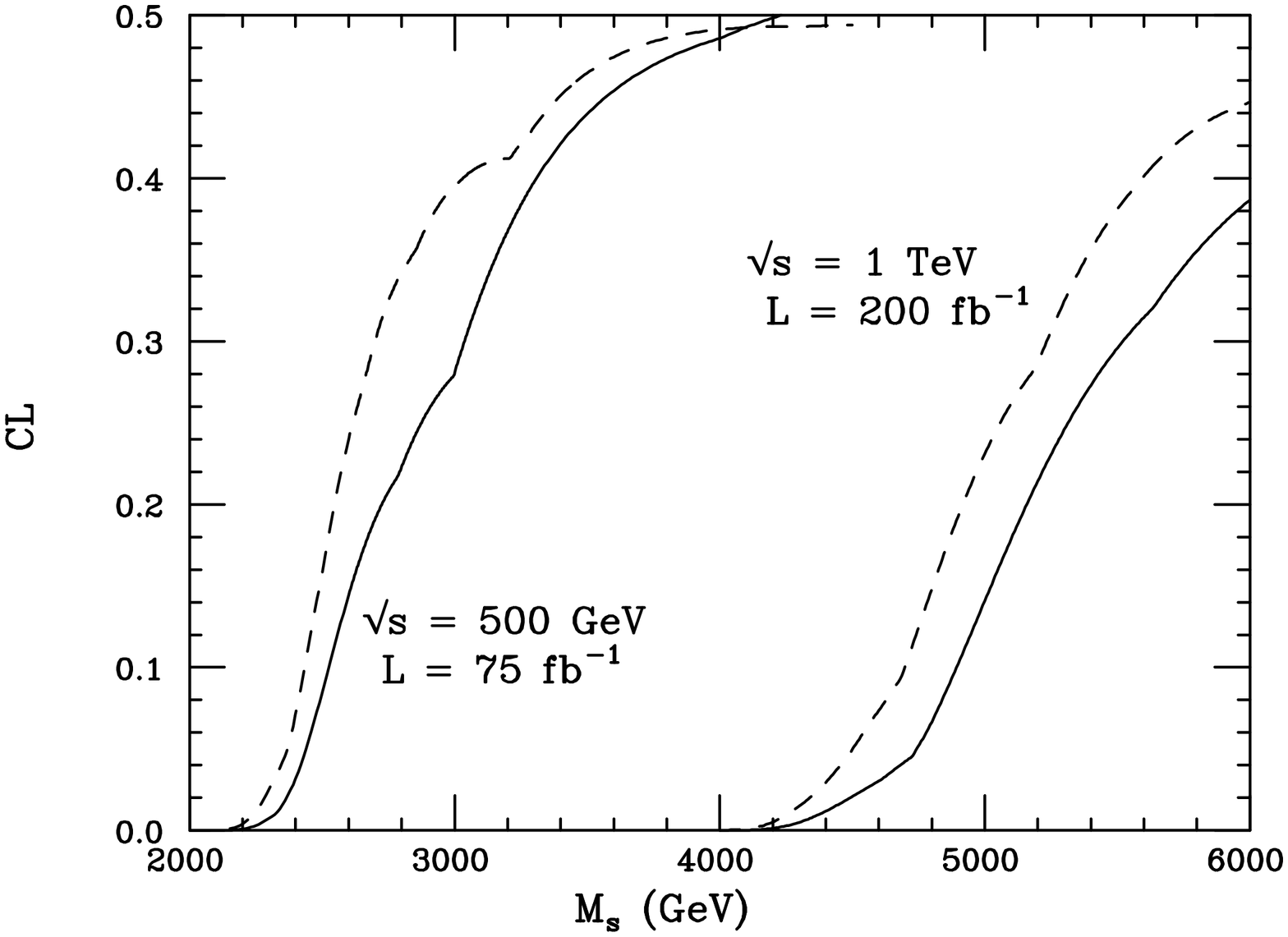,height=8.cm,width=8cm,angle=-90}}
\vspace*{-0.7cm}
\caption{Sensitivity of the LC to low-scale quantum gravity, from 
Ref.~\protect\cite{Hewett}. Left: 95\% confidence level exclusion reach as a
function of integrated luminosity for $\sqrt{s}=500$~GeV and 1~TeV.
 Right:Confidence level of fitting low-scale quantum gravity ``data'' with a
given $M_s$ to the hypothesis of spin-1 exchange, demonstrating the ability of
the LC to distinguish spin-2 from spin-1 exchange for $M_s$ up to about 
$5\sqrt{s}$. The dashed (solid) curves correspond to $\lambda=+1(-1)$.}
\label{fig:hewett}
\end{figure}

One way to search for the effect of these large extra dimensions at the LC 
is through the effect of graviton exchange on fermion pair 
production\cite{Hewett}.
This process is extremely well-understood theoretically and is a sensitive
probe of many types of new physics, including $Z^{\prime}$'s, compositeness, 
and technicolor. Graviton exchange turns out to leave the total cross
section and integrated left-right asymmetry unchanged, but modifies the
angular distributions in a way that depends on a single parameter, $\lambda/M_s^4$.
Here $\lambda$ is a dimensionless parameter of order one (but of either sign) 
that depends on model-dependent physics above $M_s$. A fit to the angular
distribution of $\epem\rightarrow \ell^+\ell^-, b\bar{b}$, and $c\bar{c}$ gives
the exclusion reach shown in Figure~\ref{fig:hewett}(a). A 1~TeV LC with 
200~fb$^{-1}$ can exclude $M_s$ up to 6.6 TeV, similar to the 6.0~TeV achievable
with the LHC in 100~fb$^{-1}$ using $\epem$ and $\mu^+\mu^-$ final states only. 
However, the LHC may have difficulty distinguishing
a graviton signal from some other new physics process, such as a $Z^{\prime}$.
At the LC, the polarized beams and the ability to observe $b\bar{b}$ and $c\bar{c}$
final states allow a clear separation between spin-1 and spin-2 exchange 
for $M_s$ up to about $5\sqrt{s}$, as shown in Figure~\ref{fig:hewett}(b).

\section*{Conclusions}
TeV-scale $e^+e^-$ linear colliders offer complementary access to the 
physics of electroweak symmetry breaking that will be explored initially
by the LHC. Assuming that both the NLC and TESLA designs prove technologically
(and financially) feasible, the choice of which one to build may 
depend on the relative importance of high luminosity for the highest precision 
measurements at lower energies (TESLA), versus upgradability to 1-1.5 TeV for 
exploratory physics and possible fuller elucidation of the SUSY spectrum.
More advanced accelerator designs, such as the two-beam CLIC\cite{CLIC} design, 
may open the path to even higher energies in coming decades, ensuring a 
vibrant future for \epem~physics in the post-LHC era.

\section*{Acknowledgements}
I would like to thank the organizers for a stimulating and enjoyable 
conference in a lovely setting. This work is supported in part by 
DOE contract number DE-FG02-95ER40899 and by NSF CAREER award PHY-9818097.

\end{document}